# A METHOD OF PREDICTING POWDER FLOWABILITY
# FOR SELECTIVE LASER SINTERING

D. Sassaman*[1], T. Phillips*[1], J. Beaman*, C. Milroy‡, and M. Ide†

*Department of Mechanical Engineering, The University of Texas at Austin, Austin, TX, 78712
‡Texas Research Institute Austin, TX 78746
†ExxonMobil Research and Development Company, Annandale, NJ 08801

## Abstract

This work investigates a method for pre-screening material systems for Selective Laser Sintering (SLS) using a combination of Revolution Powder Analysis (RPA) and machine learning. To develop this method, nylon was mixed with alumina or carbon fibers in different wt.% to form material systems with varying flowability. The materials were measured in a custom RPA device and the results compared with as-spread layer density and surface roughness. Machine learning was used to attempt classification of all powders for each method. Ultimately, it was found that the RPA method is able to reliably classify powders based on their flowability, but as-spread layer density and surface roughness were not able to be classified.

## Introduction

As additive manufacturing (AM) becomes more widespread, an emphasis has been placed on speeding up development of new materials that offer functionality beyond what is currently available. This work will focus on one AM process, Selective laser sintering (SLS), which utilizes a powder feedstock to produce polymeric components. One method of producing complex materials with SLS is to employ an indirect approach where the feedstock is a mixture of melting and non-melting components. The melting component, typically a polymer, acts as a binder for the non-melting components, which add functionality to the composite. Typically, these composite materials undergo post-processing steps to achieve their desired properties after their geometric properties have been imparted by the SLS system. Developing new materials for indirect SLS is especially challenging due to interactions between the constituent materials that affect their performance in SLS.

Currently, a trial-and-error approach is taken when developing new indirect SLS materials. Fully testing a new material in a commercial SLS system requires multiple kilograms of material, a requirement that can be costly and time intensive. Therefore, a pre-screening process that can quickly determine a material's suitability for SLS using small volumes of material can improve the development process. In this paper, relative density, powder flow, and compaction characteristics are considered as screening criteria.

This work began with the hypothesis that there exists a simple metric to determine a powder's flowability and compaction characteristics prior to the SLS process. In other words, there is a link between this *a priori* metric and some physical characteristics of the as-spread powder in the SLS system. It turns out that this link is difficult to determine and the current methods of pre-screening SLS material do not explain the differences in manufacturability of different materials.

---

[1] Contributed equally to this project

## Background

A pre-qualifying (or screening) step is often employed in powder-based manufacturing processes. The purpose of this measurement is to reduce amount of trial and error, and expedite the process for determining if a powder, or mixture thereof, will work for a given application. Applications which employ such a step are broad, and include storage in silos, pharmaceuticals, and packed-bed reactors [1]. The screening process, an empirical step, is often required because theoretical modeling of the powder behavior is complex. Unfortunately a *universal* qualifying methodology does not exist and is impractical; it must reflect the conditions of the powder in the process under consideration [1]. The screening processes is especially useful if many possible powder systems have been selected based on application requirements, but have not yet been proven viable candidates for a given manufacturing process [2].

A multitude of screening metrics and methods for powders in SLS have been studied in the literature [3]–[6]. Amado *et al*. proposed that powders for SLS have intrinsic and non-intrinsic properties [3]. The intrinsic properties, such as melt temperature, are related to the chemical and physiochemical properties of the powder. Non-intrinsic properties, such as particle size, are more closely related to the pre-processing and production steps. In the case of SLS, non-intrinsic properties are those typically associated with "dispersion, packing, and homogeneity" of each powder layer [7]. The intrinsic properties of powders for SLS play a larger role during laser irradiation. Powder properties can be further categorized into static and dynamic. Static properties are those not altered by physical movement and/or do not change during the SLS process. Dynamic properties are either altered during the SLS process, *or* experience physical movement.

|  | Intrinsic | Extrinsic |
|---|---|---|
| Static | Melt Temp. | Particle Size & Shape Bulk Density |
| Dynamic | Melt Viscosity | Powder Flowability Packing Efficiency |

*Figure 1: Material property categories for powders in SLS, proposed by Amado et al [3]*

Powder layer homogeneity and compaction are directly related to the surface quality and density of the final parts produced with SLS, and research suggests that *powder flowability* and *compressibility* are the key metrics for predicting this [1]. Unfortunately, a universal method of measuring these does not exist. On top of that, there is no clear correlation *between* methods, making comparison difficult.

A common measurement of *powder compressibility* is the Hausner Ratio (Eqn. (1), which relates the bulk to tapped density of the powder, and is explained further in ASTM D7181 [8]. However, the Hausner Ratio does not closely resemble powder spreading in SLS, so some suggest that it may not be the best measurement [6]. Additionally, the tapped density can be difficult to perform with repeatability (it is strongly affected by the frequency and amplitude of tapping) and often has a high standard deviation [9], [10].

$$HR = \frac{\rho_{tapped}}{\rho_{bulk}} \tag{1}$$

Van den Eynde *et al* developed a powder flowability testing device which aims to more closely resemble the stress states during powder spreading in SLS (Figure 11 in Appendix A). They used a *modified* Hausner ratio, called Packing Factor (Eqn. (2), which takes into account powder bed density [6], [11]. However, Packing Factor still uses tap density, an unreliable measurement. Verbelen *et al*. further utilized this device to analyze a number of commercial nylon powders for SLS [5].

$$PF = \frac{\rho_{layer}}{\rho_{tap}} \tag{2}$$

Amado *et al*. proposed the use of a revolution powder analyzer (RPA), to complement existing measurement techniques, as it closely resembles the dynamic powder spreading process during SLS (Figure 11). Schmid and Siegelmeier *et al*. reviewed the various methods of measuring powder flowability, cohesion, and packing efficiency [4], and found the RPA to have the lowest standard deviation, and perhaps, be the best representation of spreading in SLS.

## **Methods**

This work combines existing pre-qualifying metrics with machine learning (ML) to investigate if a simple material-agnostic method exists to predict powder flow and compaction. Additionally, other pre-qualifying metrics from literature are evaluated and compared to the RPA approach.

The RPA method consisted of putting 20 ml of material into a glass jar 55mm diameter by 27mm height. The jar was placed on a rolling mill and spun at 37 rpm. Side illumination was used to enhance contrast at the powder-air interface and a video camera was used to capture the data at 30 frames per second for 60 seconds. The video data was processed in imageJ to extract the powder-air interfaces. Figure 2(A) presents a still image from the video data and Figure 2(B) shows the powder-air interface that is extracted.

Flowability metrics were extracted from the RPA video data by fitting two lines to each powder-air interface, the left line containing the avalanching powder and the right line containing the fluidized powder as seen in Figure 2(C). The fit data was broken down into 30-frame segments and fit statistics were extracted from each segment. A total of 60 segments (1800 frames) were taken for each material. The key statistics measured from each segment were the mean and standard deviation of the slopes as well as the average root-mean-square error (RMSE) values. These 6 metrics were used to define the *flowability* of each data segment.

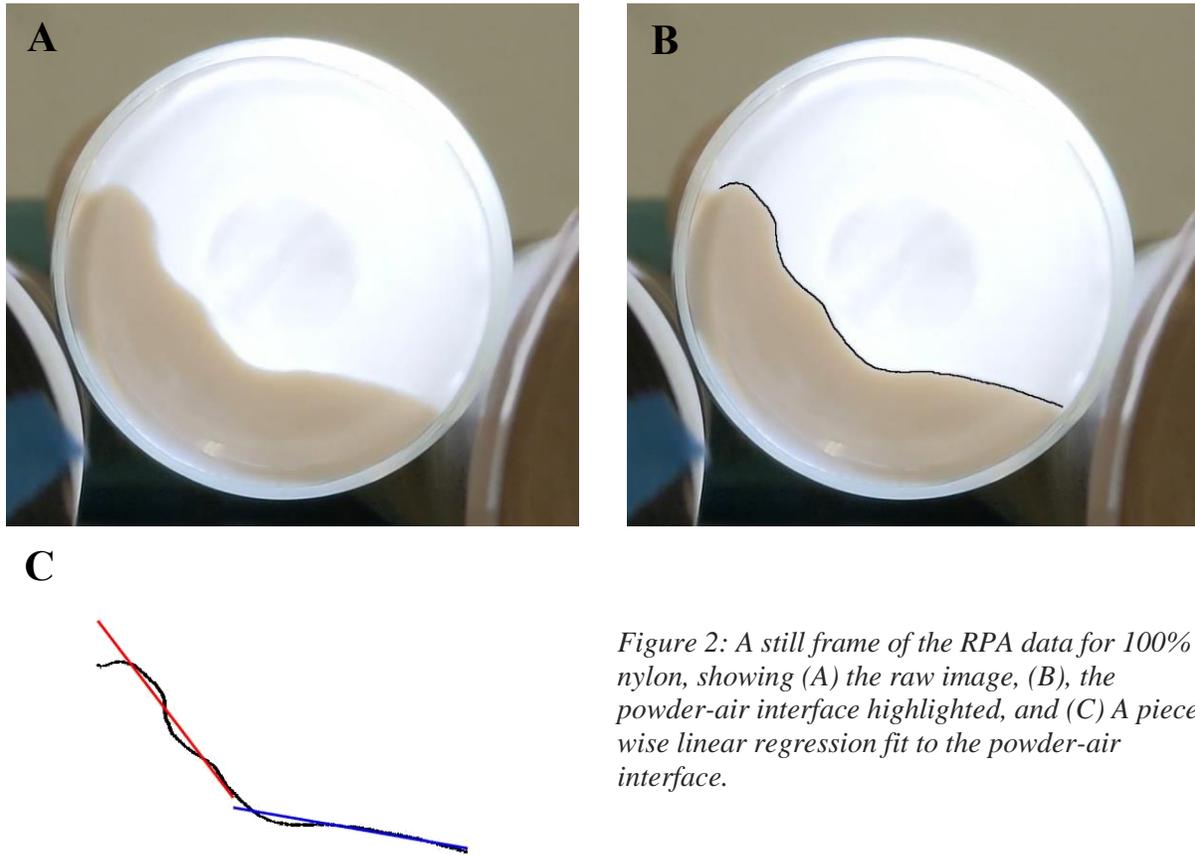

*Figure 2: A still frame of the RPA data for 100% nylon, showing (A) the raw image, (B), the powder-air interface highlighted, and (C) A piecewise linear regression fit to the powder-air interface.*

The 6 key metrics extracted from each RPA data segment were compiled into feature vectors and used in a quadratic support vector machine (SVM) to determine if the powders could be differentiated, as depicted in Figure 3. Classification was performed in Matlab using 5-fold cross-validation.

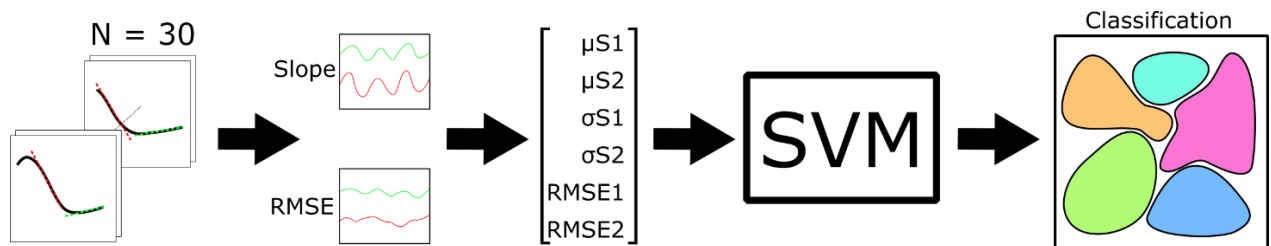

*Figure 3: Extraction of flowability metrics from RPA data. From left to right: the powder-air interface is recorded for a segment of 30 consecutive frames. From each segment, piecewise linear regression statistics are distilled into a feature vector. The feature vectors are used to train a support vector machine capable of classifying the powders.*

The other pre-qualifying metrics investigated were: Hausner ratio (bulk density and tapped density), packing factor (spread density and tapped density), pressed density, absolute density, and spread-layer surface roughness. Bulk density was measured by pouring 70mL of powder through a funnel (30°, 4.7mm opening) into a 100mL graduated cylinder and measuring the powder's mass. Tap density was determined by taking the powder-containing graduated cylinder from above and tapping on a hard surface from a height of ~15mm for 2 minutes at roughly 2Hz. Pellets of each powder system were produced in a 1" die press using 1000±10 kgf, and their outer dimensions measured using a dial caliper. All masses were obtained using a balance with resolution of ±0.001gram. True (absolute) density was measured using helium pycnometry (Quantachrome Ultrapyc 1200e, 20psi purge, 3-run avg.). Powder bed density was determined by spreading a 100μm layer of powder onto a known volume using the counter-rotating roller in a 3D-Systems HiQ. The baseplate of known volume was placed on top of balance with resolution of ±0.1gram. A process schematic for these metrics is shown in Figure 4. Results for bulk, tapped, true, and pellet density are provided in **Error! Reference source not found.**.

The as-spread surface roughness was measured using a Keyence VHX7000 optical profilometer[2]. Each powder was spread into a 500μm deep pocket manually using a doctor blade (round-edge, 45° blade angle); six replicates of each powder were performed. Five surface roughness parameters were calculated using the Keyence software for the entire scanned area (~57mm$^2$); mean height deviation ($S_a$), maximum peak-trough height ($S_z$), RMSE of height deviation ($S_q$), mean skewness ($S_{sk}$), and mean kurtosis ($S_{ku}$). The surface roughness parameters were compiled into feature vectors and used in a quadratic SVM to determine if the powders could be differentiated. Classification was performed in Matlab using 5-fold cross-validation, similar to Figure 3.

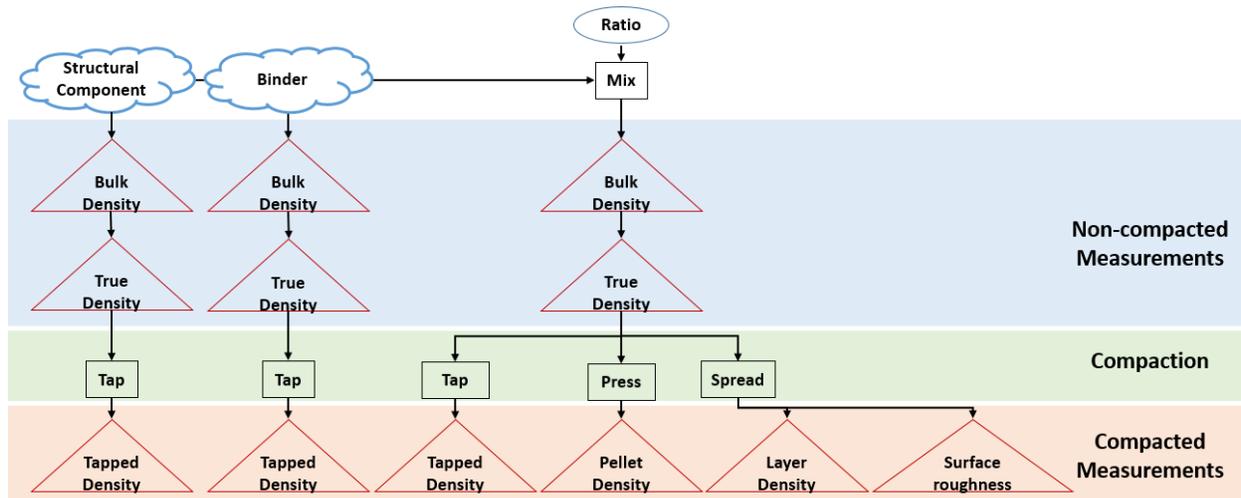

*Figure 4: Pre-qualifying metrics investigated in this study*

---

[2] 10μm step size, 200X magnification, 5 x 5 stitching (57mm$^2$ total area)

Two powder systems were investigated in this study: aluminum oxide + nylon, and carbon fiber + nylon. The alumina/nylon mixture was adapted from Deckers, Shahzad, *et al.* [14] with slight modification[3]. The carbon fiber/nylon[4] mixture was designed as a proxy for carbon fiber-reinforced structural polymers. An array of powder mixtures was produced, each mixture with different weight percentages of polymer binder and structural material. For the remainder of this document, the names of the powder mixtures will be encoded as "wt.% structural component - wt.% nylon". For example, 10C-90N indicates 10 wt.% carbon fiber and 90 wt.% nylon, and 80A-20N indicates 80 wt.% alumina, 20 wt.% nylon.

## **Results**

The RPA data show that there is a clear difference in flowability among the materials tested. Heat maps of the powder-air interface for two materials, 100N and 10C-90N, are given in *Figure 5*. The color scale represents percent likelihood that the interface passes through that pixel throughout the 60 seconds of RPA data. In both data sets, the material is more stochastic near the jar edges and more consistent near the center of the interface; however, the 100N material displays more uniformity than the 10C-90N material.

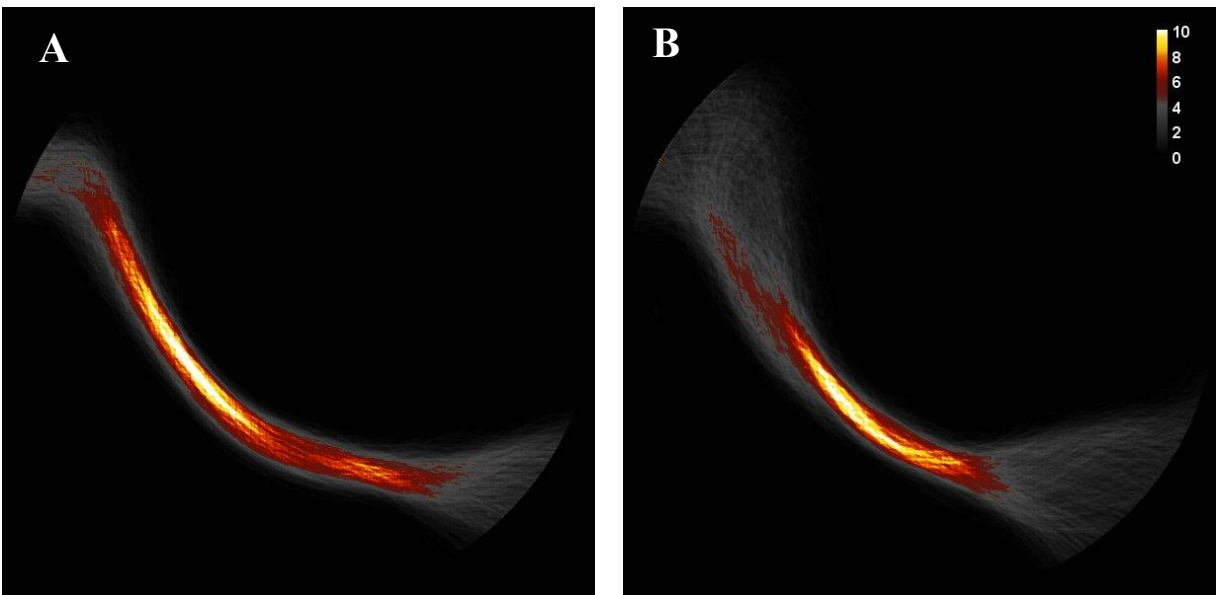

*Figure 5: Powder-air interface heat map for (A) 100% nylon and (B) 90% nylon with 10% carbon fiber. The color scale represents the percent likelihood that the interface passes through a given pixel.*

The box and whisker plots, Figure 6, show that the different materials have different distributions for the 6 features, indicating that the material flow properties, as measured with RPA, are unique for each material.

---

[3] Dry-mixed in high-shear blender (Chulux QF-TB159008) for 10 minutes. Alumina: (Almatis A16 SG, $d_{50}$=0.5µm), nylon: PA12 (ALM PA650 $d_{50}$=55µm ). Mixture  sieved <250µm
[4] Dry-mixed in high-shear blender (Chulux QF-TB159008) for 10 minutes. Nylon: PA12 (ALM PA650 $d_{50}$=55µm ), carbon fibers: (Zoltek PX-30 avg. length=100µm, avg. width=7 µm).

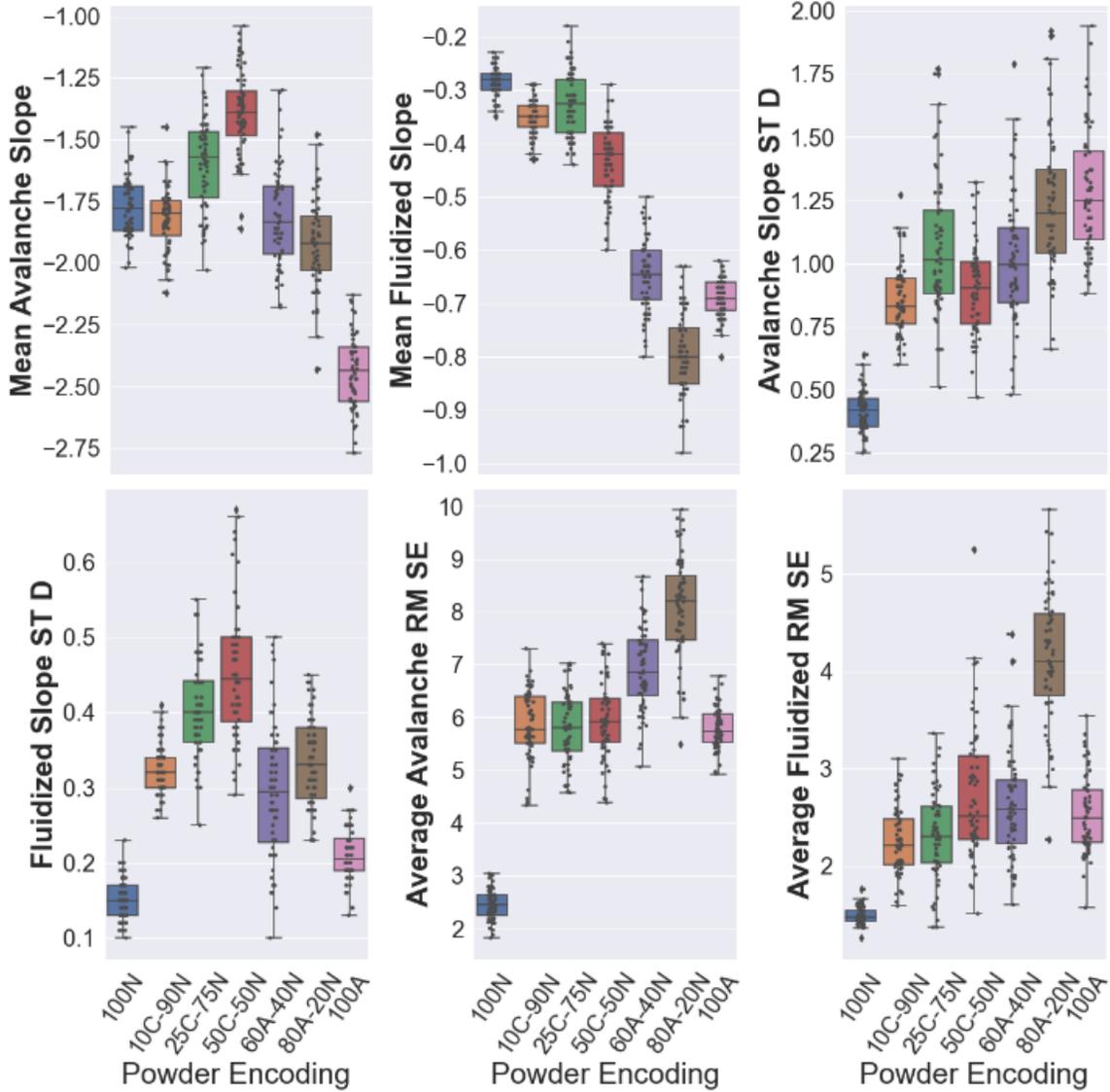

*Figure 6: Box and whisker plots for the RPA flowability metrics.*

Quadratic SVM results indicate that each video segment could be classified with an accuracy of 93.1%. The misclassification rate is given in the confusion matrix shown in Figure 7. The confusion matrix shows there is no misclassification based on material additive, i.e. no material with alumina additive was misclassified as one with carbon fibers and vice versa. The high classification accuracy suggests that all the materials tested inherently have different flow characteristics; however, the classification does not suggest anything about quality of flow, only that different materials can reliable be differentiated.

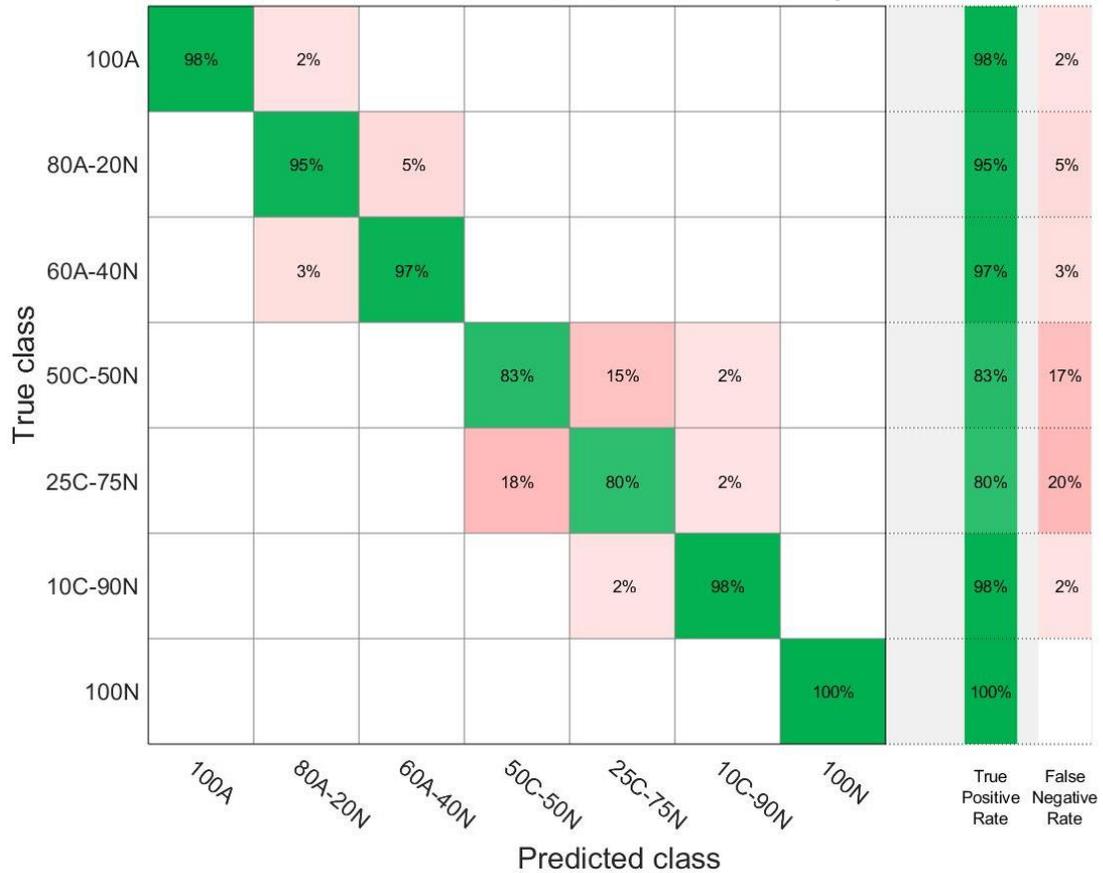

*Figure 7: Confusion matrix for classifying powders using a quadratic support vector machine.*

Despite visual indications that each powder provided a different as-spread surface roughness (Figure 8), attempts to classify the powders using a quadratic SVM were unsuccessful (<62.5% accuracy). Figure 9 shows a box and whisker plot for the surface roughness parameters, showing that many metrics are overlapping for the different materials. For example, a $S_z$ measurement of 0.4mm could be from any of the materials tested. The overlapping nature of these measurements helps explain why classification was unsuccessful.

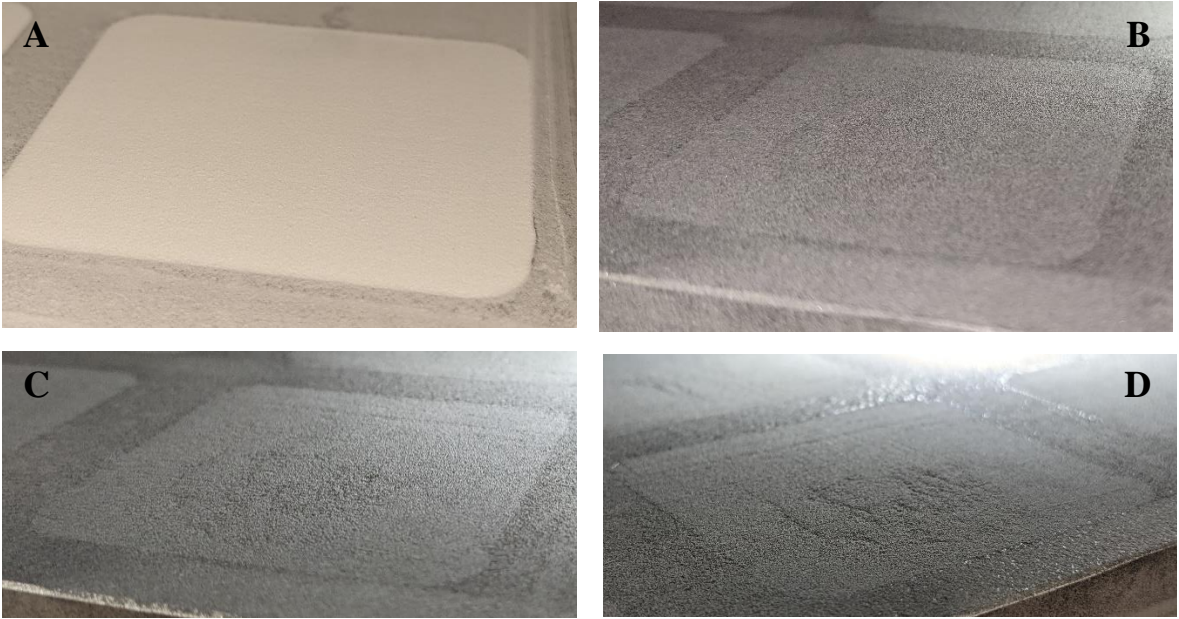

*Figure 8: Surface roughness of various nylon/CF powders.*
*A) 100 wt.% nylon, B) 10wt.% CF / 90wt.% nylon*
*C) 25wt.% CF / 75wt.% nylon, D) 50wt.% CF / 50wt.% nylon*

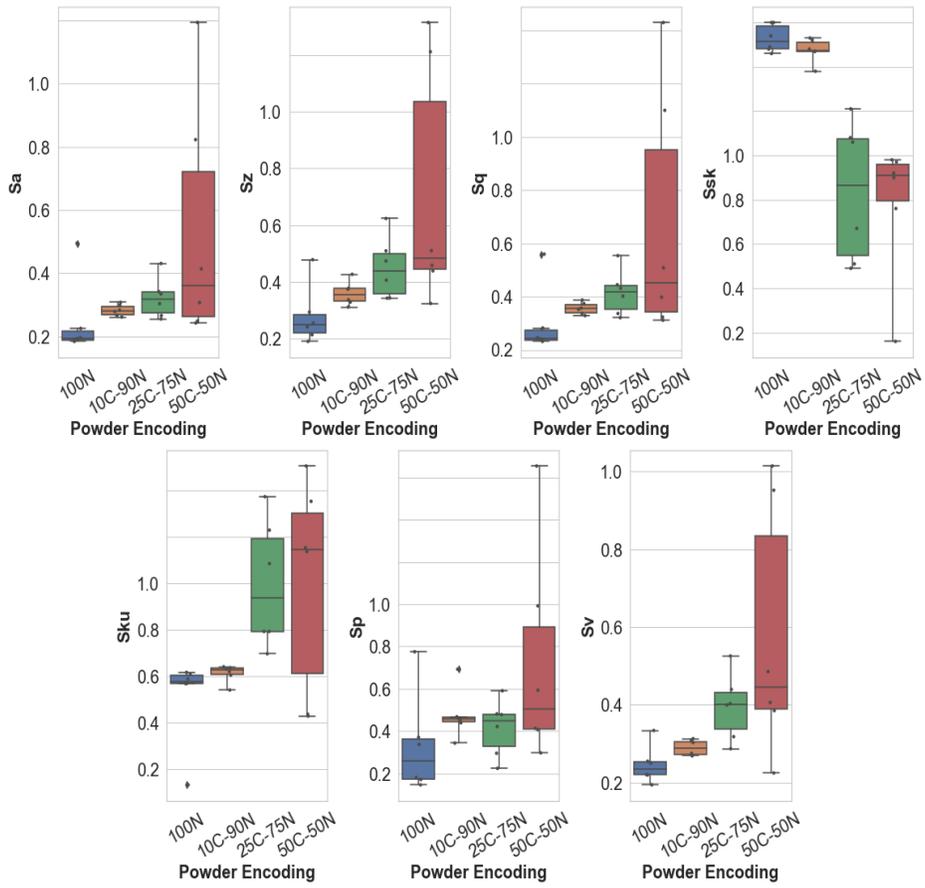

*Figure 9: Surface roughness parameters for CF/nylon powders*

The most common flowability and compaction metrics in literature were also investigated: Hausner ratio (bulk density and tapped density), packing factor (spread density and tapped density), and pressed density. These results are shown in Figure 10 and

Table 1. Accurate powder classification was not possible with any of these metrics, indicating that, despite different flow and compaction characteristics for each powder, these methods do not have a fine-enough resolution to be used for reliable pre-qualification for SLS.

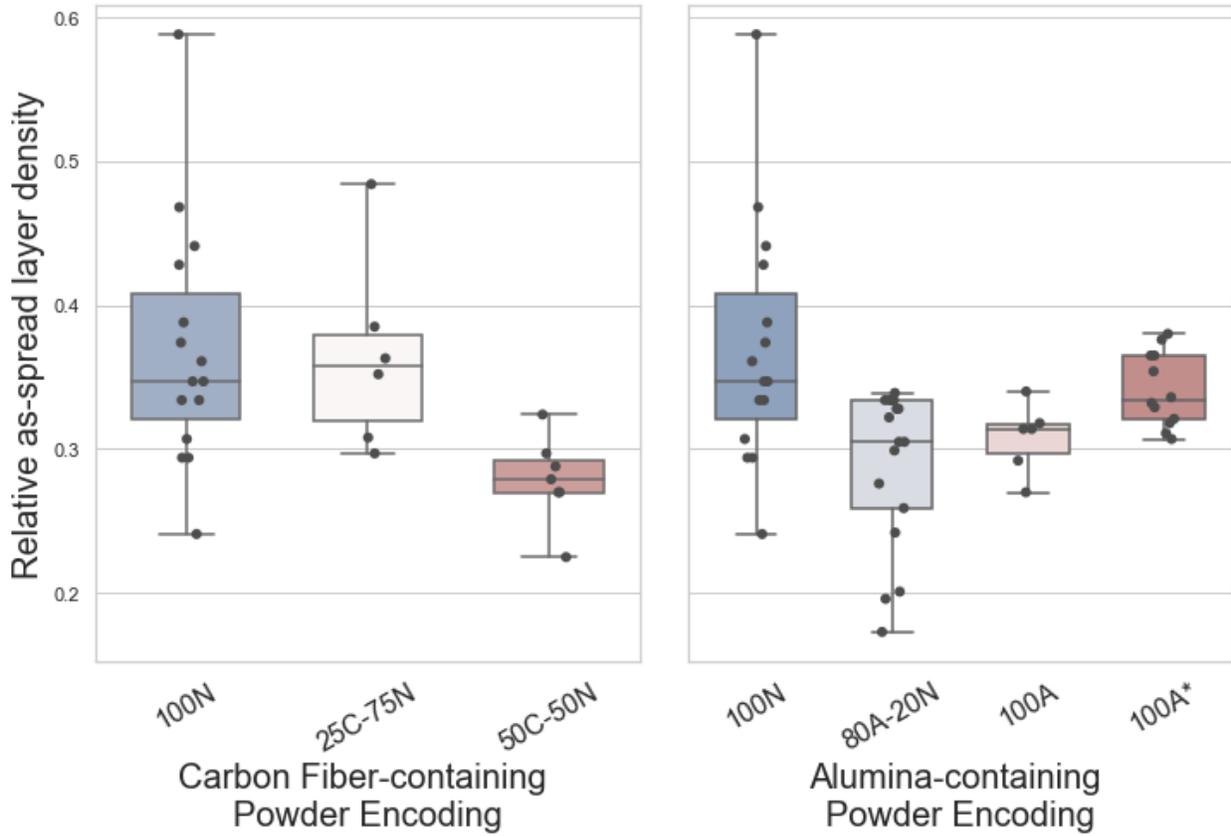

*Figure 10: As-spread layer densities for carbon- and alumina-containing powders.*
*Relative density calculated by dividing as-spread density by true density obtained by helium pycnometry*
*Note: 100A\* is unsieved alumina, 100A is sieved <250μm*

Table 1: Densities for powders in this study. The ± values correspond to one standard deviation.

| Powder | Theoretical[5] & Literature Values | | | | Measured Values | | | | |
|---|---|---|---|---|---|---|---|---|---|
| | True Density [g/cc] | Bulk Density [g/cc] | Tapped Density [g/cc] | Hausner Ratio | True Density [g/cc] | Bulk Density [g/cc] | Tapped Density [g/cc] | Hausner Ratio | Pellet Density [g/cc] |
| 100N | 1.02 | 0.46 | 0.50[6] | (1.15-1.23)**Error! Bookmark not defined..Error! Bookmark not defined.** | 1.06 | 0.47 ±0.02 | 0.52 ±0.02 | 1.11 ±0.04 | 0.65 |
| 60A-40N | 2.75 | 0.83 | 0.99 | 1.19 | 1.94 | 0.64 ±0.01 | 0.78 ±0.05 | 1.23 ±0.06 | 1.12 ±0.02 |
| 80A-20N | 3.32 | 0.96 | 1.16 | 1.21 | 2.47 | 0.73 ±0.01 | 0.91 ±0.05 | 1.25 ±0.06 | 1.45 ±0.01 |
| 90A-10N | 3.61 | 1.02 | 1.24 | 1.22 | 3.29 | 0.82 ±0.02 | 1.07 ±0.10 | 1.31 ±0.12 | 1.68 ±0.03 |
| 100A | 3.90 | 1.08 | 1.32[7] | 1.22, 1.45[8] | 3.88 | 1.06 ±0.04 | 1.29 ±0.06 | 1.21 ±0.10 | 2.02 |
| 100N | 1.02 | 0.46 | 0.50[9] | (1.15-1.23)**Error! Bookmark not defined..Error! Bookmark not defined.** | 1.06 | 0.47 ±0.02 | 0.52 ±0.02 | 1.11 ±0.04 | 0.65 |
| 10C-90N | N/A | N/A | N/A | N/A | 1.18 | 0.39 | 0.47 | 1.21 | 0.63 |
| 25C-75N | N/A | N/A | N/A | N/A | 1.29 | N/A | N/A | N/A | N/A |
| 50C-50N | N/A | N/A | N/A | N/A | 1.57 | 0.48 | 0.61 | 1.28 | 0.68 |

## **Conclusion**

The RPA method was shown to be sensitive enough for pre-qualification of powder for SLS, reliably classifying materials using a very low volume of material. However, a link from the RPA metrics to a physical quantity such as as-spread layer density or surface roughness was not possible. Without linking the RPA data to a physical indicator of powder flowability, the results will not be able to predict a material's suitability for SLS. The RPA method presented here does show promise as an SLS material screening process and could be used in the future with a subjective classification of suitable as-spread layers, or possibly, an objective classification using alternative imaging techniques of the powder surface.

---

[5] Weighted sum. E.g. $\rho_{mixed} = \rho_1 * wt_1 + \rho_2 * wt_2$
[6] Van den Eynde et al. 2015 [6]
[7] German and Bose 1997 [12]
[8] De Oliveira et al. 2019 [7]
[9] Van den Eynde et al. 2015 [6]

## Appendix A: Pre-qualifying methods from literature

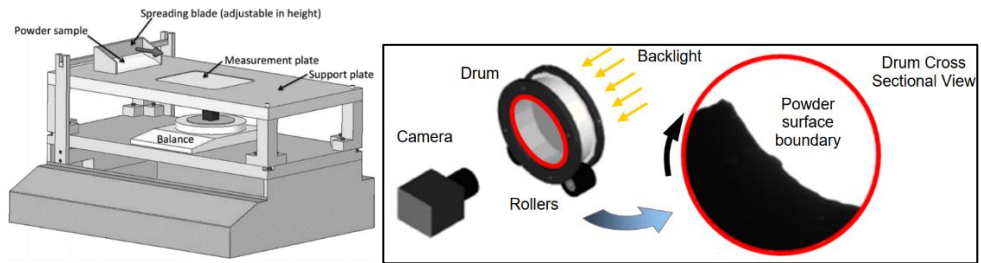

*Figure 11: Left) Spread layer density measurement by Van den Eynde et al. [6]*
*Right) Schematic of revolution powder analyzer from Amado et al.[3]*